# How to restart? An agent-based simulation model towards the definition of strategies for COVID-19 "second phase" in public buildings

Marco D'Orazio[1], Gabriele Bernardini[1], Enrico Quagliarini[1]

[1] DICEA Dept, Università Politecnica delle Marche, via di Brecce Bianche 60131 Ancona, phone: +39 071 220 4246, fax: +39 071 220 4582, mail: m.dorazio@staff.univpm.it

**Abstract.**
Restarting public buildings activites in the "second phase" of COVID-19 emergency should be supported by operational measures to avoid a second virus spreading. Buildings hosting the continuous presence of the same users and significant overcrowd conditions over space/time (e.g. large offices, universities) are critical scenarios due to the prolonged contact with infectors. Beside individual's risk-mitigation strategies performed (facial masks), stakeholders should promote additional strategies, i.e. occupants' load limitation (towards "social distancing") and access control. Simulators could support the measures effectiveness evaluation. This work provides an Agent-Based Model to estimate the virus spreading in the closed built environment. The model adopts a probabilistic approach to jointly simulate occupants' movement and virus transmission according to proximity-based and exposure-time-based rules proposed by international health organizations. Scenarios can be defined in terms of building occupancy, mitigation strategies and virus-related aspects. The model is calibrated on experimental data ("Diamond Princess" cruise) and then applied to a relevant case-study (a part of a university campus). Results demonstrate the model capabilities. Concerning the case-study, adopting facial masks seems to be a paramount strategy to reduce virus spreading in each initial condition, by maintaining an acceptable infected people's number. The building capacity limitation could support such measure by potentially moving from FFPk masks to surgical masks use by occupants (thus improving users' comfort issues). A preliminary model to combine acceptable mask filters-occupants' density combination is proposed. The model could be modified to consider other recurring scenarios in other public buildings (e.g. tourist facilities, cultural buildings).

## 1. Introduction

The "second phase" of the COVID-19 emergency should face the possibility to gradually remove containment measures adopted in the lockdown ("first phase") phase, to restart activities thus gradually ensuring restoring pre-emergency social and economic life of the Countries. "Moving from containment to mitigation" will be one of the fundamental elements for the definition of new risk-reduction strategies, to maintain an acceptable level of COVID19 cases during the time, avoiding secondary widespread spreading conditions (Prem et al., 2020; Wilder-Smith et al., 2020). In this context, particular attention should be posed in organizing the restart of activities in public buildings like offices, schools, university and other workplaces. In such closed environment, the presence of the same users (e.g. workers, students) and the possibility of significant overcrowd conditions over space and time could amplify the possibility of prolonged contact with infected occupants (Adams et al., 2016; Gao et al., 2016; Zhang et al., 2018). In fact, the users can be daily and weekly exposed to the infections within the same rooms, during the whole period, with the possibility to stay at a limited distance and in a closed environment for a long time due to the performed activities. Such environmental factors could seriously increase the possibility of a user to be infected by the novel coronavirus SARS-CoV-2 (Yang et al., 2020), as for other similar viruses (Adams et al., 2016; Gao et al., 2016; Pica and Bouvier, 2012; Prussin et al., 2020; Wilder-Smith et al., 2020). Real-world cases of closed environment where the virus spread underline such possible effects on the occupants (see e.g. the Diamond Princess cruise emergency) (Fang et al., 2020; Mizumoto and Chowell, 2020).

The current rules for the virus spreading released by the health organizations are based on proximity-based point of view (Bourouiba, 2020; Yang et al., 2020). Within the number of conditions evidenced by the national and international health organizations[1], a probable case is a person: (a) having "face-to-face contact with a COVID-19 case within 2m and > 15 minutes" in any kind of environment, whose conditions include any direct contact with a confirmed infected case (including those related to physical contact, e.g. shaking hands, contacts with the infected case's secretions); (b) remaining "in a closed environment (e.g. classroom, meeting room, hospital waiting room, etc.) with a COVID-19 case for 15 minutes or more and at a distance of less than 2 m", whose conditions also include any indirect contact according to a proximity-based standpoint. Such a proximity-based approach is in line with general studies on the mechanisms

---

[1] e.g.: http://www.salute.gov.it/portale/nuovocoronavirus/dettaglioFaqNuovoCoronavirus.jsp?lingua=italiano&id=228 ; https://www.ecdc.europa.eu/en/case-definition-and-european-surveillance-human-infection-novel-coronavirus-2019-ncov (last access: 8/4/2020)

leading to airborne diseases transmission (Chen et al., 2020; Pica and Bouvier, 2012; Prussin et al., 2020). At the same time, a higher probability to be infected can be associated to the absence of individual's protection-based measures (respiratory protective devices, such as facial masks) (Bourouiba, 2020; Chen et al., 2020; Murray et al., 2020; Zhai, 2020). Basing on experimental data also related to closed environments, previous studies tried to trace the general virus characterization in terms of: transmittal potential (Mizumoto and Chowell, 2020); incubation period and timing to symptoms onset (e.g. fever) (Lauer et al., 2020; Wilder-Smith et al., 2020); asymptomatic ratio (Mizumoto et al., 2020). Finally, the transmittal potential seems to increase in case of infectors' symptoms onset, as for other viruses (Wilder-Smith et al., 2020; Yang et al., 2020).

In view of the above, three main strategies could be activated to mitigate the virus spreading (Barbieri and Darnis, 2020; Chan et al., 2009; "Face ID firms battle Covid-19 as users shun fingerprinting," 2020; Fang et al., 2020; Howard et al., 2020; Prem et al., 2020; Servick, 2020; Yang et al., 2020; Zhai, 2020)[1]: (a) social distancing solutions, which essentially lead to the closure of public buildings and the reduction of mixing in communities during the lockdown phase of COVID-19 emergency; (b) the use of respiratory protective devices (facial masks); (c) control of COVID-19 infectors, starting from the detection of the symptoms (e.g. fever) to isolate the possible infected cases, up to the tracing of infected cases motion to limit their contact with susceptible people.

The combination of such strategies will be fundamental to remove the lockdown in public buildings (Yang et al., 2020). Nevertheless, it is necessary to understand the effective impact of each measure (and of their combination) to effectively set up acceptable solutions from the perspective of both stakeholders (e.g. which solutions can be easily implemented with an effective impact on the activity?) and the final users (e.g. which solutions will lead to the possibility to move towards the "normal" fruition of the spaces?) of public buildings. To this aim, simulation tools could be useful in predicting how different mitigation solutions could affect the virus spreading, as also remarked by previous studies on airborne diseases mitigation (Emmerich et al., 2013; Gao et al., 2016; Laskowski et al., 2011; Saari et al., 2006; Zhang et al., 2018). The importance of such tools has been evidenced in many different cases concerning individuals' safety in the Built Environment, when risks for users are also related to individuals' behaviours, including motion issues, such as those related to emergency safety and evacuation (Bernardini et al., 2017; D'Orazio et al., 2014; Zheng et al., 2009).

Concerning the COVID-19 emergency, different tools have been developed according to experimental data to derive general rules for the contagion spreading at a wide-scale (e.g. national scale) and the impact of risk-mitigation strategies on the number of infected people (Fanelli and Piazza, 2020; Lopez and Rodo, 2020; Prem et al., 2020), while limited efforts have been performed towards the definition of models in a closed environment (Fang et al., 2020; Mizumoto and Chowell, 2020). Nevertheless, as suggested by previous works (including simulation-based ones) taking into account both the public building occupancy and the movement of the same occupants in a wider scale context (e.g. city) (Gao et al., 2016; Prem et al., 2020; Zhang et al., 2018), the permanence of people in the workplaces (such offices) and school will lead to one of the more significant contributions in the contagion spreading, also because of the permanence timing and of the crowd level and social contacts ways (e.g. possibility to have close contacts/contacts within a radius of 2m).

This study provides a simulation model to estimate the virus spreading in the closed built environment by adopting a proximity-based approach according to the aforementioned recognized rules for contagion. By this way, all direct and indirect contagion effects between individuals placed at a close distance can be simulated. The model adopts a probabilistic approach to jointly simulate the building fruition by the users (motion rules, occupancy during the time of some areas to perform activities) and the virus transmission between them. An agent-based modelling approach is chosen to define the specific simulation rules at each involved agent (Banos et al., 2015; Laskowski et al., 2011), which can be described also in terms of spreading-affecting additional features (i.e. wearing a facial mask, being at a particular moment of the incubation period, being asymptomatic or not) (Fang et al., 2020; Mizumoto and Chowell, 2020). In this view, the stochastic approach also enables to create different input scenarios and manage differences in the spreading trend inside the building. Meanwhile, the agent-based model allows considering microscopic rules for the simulated individuals (D'Orazio et al., 2014), in this case consisting of the characterization of both use-of-spaces and virus-related issues (e.g. symptoms onset, use of respiratory protective devices, and so on) (Fang et al., 2020). The proposed model is calibrated according to the application to a real COVID19 emergency scenario in a closed environment, using comparison of simulation results and experimental data (Mizumoto and Chowell, 2020). Finally, the application of the calibrated model is provided for a significant case study (a university building) to evaluate the impact of different mitigation strategies on the infected people's number.

## 2. Phases, model description and methods

The work is divided into the following phases:
1) definition of the modelling approach (see Section 2.1);
2) implementation of the model within a simulation software, and calibration activities according to experimental data related to a significant closed environment (Diamond Princess case study) (see Section 2.2);
3) application to a relevant case study, by considering a sensitivity analysis-based approach, and then evaluating the impact of different spreading-mitigation strategies (i.e.: respiratory protective devices; crowd level control; infectors' access control) (see Section 2.3).

### 2.1. Modelling approach

An agent-based modelling approach is used to evaluate the spreading of COVID-19 into the closed Built Environment by considering the possible movement of the simulated agents discretely placed in the environment, to jointly considering the two main dynamics of the matter (epidemic and mobility-related) (Banos et al., 2015; Fang et al., 2020).

Concerning the epidemic-related aspects, according to national and international health organization[1], previous works on preliminary COVID-19 characterization and simulation models for virus spreading (Fang et al., 2020; Yang et al., 2020), it is assumed that the main driver is the distance between a virus carrier and the other individuals placed into the Built Environment (Adams et al., 2016; Prussin et al., 2020). This proximity-based contagion spreading approach includes all direct and indirect spreading effects between two individuals. Furthermore, a latent period for infected agents can exist (Lauer et al., 2020), called "*delay*" period.

In view of these common aspects, each simulated agent is characterized by:

- being infected or not, which can vary at each step of the simulation. When the simulation starts, there is a certain percentage of occupants that is infected inside the Built Environment. Once an individual is infected, he/she can become an *effective infector* after the initial *delay* period (latent period), which corresponds to the initial phase during which the virus load increases up to provoke secondary infections. The contagion timing is $t^*$ [simulation step]. At $t^*=0$, the agent becomes infected. Then, $I_{delay}(t^*)$ [simulation step], which is defined as the current step from the virus contagion, increases at each simulation step by a *formcoeff* ($I_{delay}(t^*)=I_{delay}(t^*-1)+formcoeff$; in case of linear relationship, *formcoeff*=1). Hence, *formcoeff* expresses the speed of the process after the moment of the contagion;
- wearing or not a respiratory protective device (in the following: *mask*). The percentage of people wearing the *mask*, as well as the *mask* filtering capacity (defined according to European standard EN 149:2001), could be decided in the setup of the model.
- being asymptomatic or not after being infected. In particular, not asymptomatic people are considered to "die" (leave/not enter the building) when the current step from the virus contagion *Istep* [simulation step] reaches the timing to symptoms (e.g. fever) onset *Ifev* [simulation step] (Lauer et al., 2020). The agent's "die" behaviour could both represent the effects of the disease for: 1) people who can be blocked by building access control system; 2) people who spontaneously leave/not enter the building due to their health conditions.

At each simulation step, the model evaluates if an *effective infector* is placed near uninfected ones, assuming a radius of 2m as base proximity distance for contagion spreading *Dvir* [m], according to current consolidated rules provided by international health organizations[1]. In these conditions, each of the *effective infector i* has a chance to infect a neighbouring uninfected agent *j* according to Equation 1, which is based on previous modelling approaches (Fang et al., 2020):

$$Pvir = \min(1, i_j \cdot t_{exp} \cdot (1 - prot_i) \cdot (1 - prot_j) \cdot p_{imm}) \quad (1)$$

where:

- $i_j$ is transmission efficiency [-] of the virus in the surrounding individual j, as in Equation 2. Simple linearity between the incubation-related inputs is assumed.

$$i = \min\left(1, \frac{Istep}{Iinc}\right) \quad (2)$$

- $t_{exp}$ [-] is the normalized exposure time of *i* in respect of *j* as in Equation 3. It depends on the exposure time $\Delta t$ [simulation step] and on the minimum contact time $t_c$ [simulation step], which can provoke the virus spreading. $t_c$ can be defined as the number of steps to simulate an equivalent time of 15 minutes. Simple linearity between the time-dependent inputs is assumed.

$$t_{exp} = \min\left(1, \frac{\Delta t}{tc}\right) \quad (3)$$

- $prot_i$ and $prot_j$ [-] are the level of protection given by the *mask filter* of the facial *masks* worn by the two agents [-]. The parameters vary from 0, which implies "no mask", to 1, which corresponds to maximum protection level. More data on the *mask filter* values adopted by this study is provided in Table 1.
- $p_{imm}$ [-] considers that "some people may be immune to the virus" (Fang et al., 2020), and it can vary from 0 to 1 (e.g., 0 corresponds to the case in which the whole population is vaccinated).

In Equation 1, the adoption of the term 1 as a comparison factor allows limiting the upper probability *Pvir* to 100%. Once *Pvir* has been calculated, a random number (varying from 0 to 1) is compared to *Pvir* to stochastically define *j* as a new infected agent when the random number is lower or equal than *Pvir*. Once the agent has become infected, he/she will change its status into *effective infector* when the *delay* value will be reached. The overall epidemiological model approach is also resumed in Figure 1.

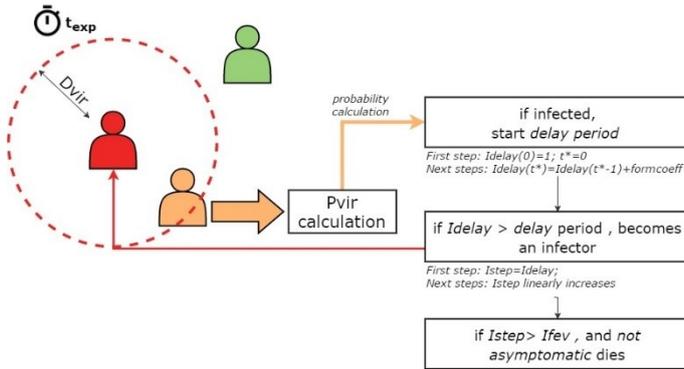

*Figure 1. General scheme for the contagion spreading: the red agent is an infector, the green agent is out of Dvir, the orange agent is a possible infected agent. The Pvir calculation also depends on the exposure timing* $t_{exp}$.

The agents are placed into the Built Environment (called "*world*"), which is represented as a unique layer whose total area depends on the gross one of the space to be simulated, to consider only the relationships between the individuals, regardless of the architectural features. The *world* is divided into patches: each patch is equal to 1m to consider a 1:1 scaled representation of the Built Environment.

Finally, this *world* is a "close" *world*: there are no births and travel into or out of the simulated population, while deaths can exist because of the virus effects on the individuals. By this way, the model can be applied in all the environment in which: new infected can essentially occur inside the *world*; infected people can remain inside until to "die". By this way, the model could be applied to all the buildings that are frequented by the same daily users, such as offices, schools, working places (Gao et al., 2016; Zhang et al., 2018): they can return home during the evening, but the following day they will share the building spaces with the other same users.

Each simulated agent can randomly move into the *world* in a certain radius of space which is defined at the simulation setup. It is considered that the movement can be performed at given time steps, to simulate the possibility that the individuals spend a certain time near the same area. In this term, each simulated agent "has the same chance of interacting with any other person within the world". Finally, the agents can be in closed contact during the use of the space.

Table 1 resumes the model parameters. In particular, $A$ [m$^2$] is the overall *world* area, which is used to estimate the area in terms of *patches*. Since the model approach considers an infection distance of about 2m and the environment shape does not alter the individuals' motion and *Pvir*, a squared area with each side equal to $A^{1/2}$ is considered into the simulator.

$N$ [pp] is the number of simulated agents into the world. As for $A$, it depends on the specific scenarios to be simulated. This factor affects the density of occupants $Docc$ [pp /m$^2$] inside the building. Lower $N$, lower the occupants' density, higher the possibility for the building stakeholders to set up social-distancing solutions inside the *world*.

Finally, the implemented model can allow agents to freely move or not in the space. In this second case (adopted in this work for the case study), after each movement, the agent returns at the previous starting position in the *world*. This essentially allows considering the agents' use of building spaces (e.g. classroom for school buildings, offices for public building and so on).

| Parameter | Unit of measure | Values range | Description |
|---|---|---|---|
| A | m$^2$, patches | >0 | the dimension of the world in which the agents move |
| N | pp | >0 | number of people in the world |
| *formcoeff* | - | >0 | speed of the virus spreading, by modifying the delay increase for each step |
| $p_{imm}$ | % | 0 to 100 | If equal to 100%, all the population is virus-resistant (e.g. due to vaccine) |
| Initial infectors % | % | 0 to 100 | how many individuals could be infectors at the starting of the simulation |
| asymptomatic ratio | % | 0 to 100 | how many individuals could be infectors without symptoms, thus not "dying" |
| average delay | simulation steps | >0 | the average delay between being an infected individual and an effective infector, due to virus replication dynamics (Lauer et al., 2020). It also depends on how many steps represent each simulation day. |

| *Iinc* | simulation step | >0 | this is the incubation time, which starts from the contagion moment to the maximum considered display symptoms time (e.g. fever onset) for all the considered population (Lauer et al., 2020). |
| --- | --- | --- | --- |
| *Ifev* | simulation step | ≥0 | this is the time from the contagion to the minimum onset of the fever (compare to *Iinc*) (Lauer et al., 2020). In this work, the value is stochastically assigned within 0 to *Iinc*. |
| $prot_i$, $prot_j$ (*mask filter*) | - | 0 to 1 | specific values can be assigned for respiratory protective devices categories for filtering half masks given by EN 149:2009 by considering maximum aerosol drops penetration percentage. Single mask characterization ranges are considered to include superior limits for each kind of mask: FFP3≥98%, 98%>FFP2≥95%; 95%>FFP1≥80%.<br>Besides, previous works tried to classify surgical mask efficiency according to the NIOSH NaCl method (Rengasamy et al., 2017), by providing an efficiency range from 54% to 88%. Finally, a no-protection limit for *mask filter* from 0 to 25% is selected to consider the non-standards protection solutions, basing on the first quartile in uniform input distribution. |
| $t_c$ | simulation step | 1 | number of steps to simulate an equivalent time of 15 minutes, according to consolidated data about indoor contagion spreading from national and international health organization[1] |
| mask wearing % | % | 0 to 100 | Percentage of people implementing the considered protection level $prot_i$ |
| movement at breaks | patches | ≥1 | maximum distance allowed during the movement for each agent, that is, the distance travelled by the simulated agent into the built environment between two different simulation steps. The probable movement distance could be derived in relation to the distance between the areas in which the individual spends time. The maximum distance could be ideally set at $A^{1/2}$ by considering a squared *A* in the patches description. |

*Table 1. Model parameters characterization*

## 2.2. Model implementation and calibration

The model described in Section 2.1 is implemented in simulation software through NetLogo (Wilensky, 1999). To ensure the application of statistical methods and the related reproducibility of scenarios in the calibration and application phases, an R script (R version 3.6.3[2]) is implemented to launch a series of simulation within the model according to previous research approaches on epidemiologic researches (Banos et al., 2015). Simulation runs were performed using NLRX package of "R statistics" programming language (Salecker et al., 2019), by defining the specific "experiment" conditions according to a series of input data

A preliminary simulation phase verified that no influence on the final result is provided by the initial distance among the simulated agents. Then, the simulation model is then calibrated by comparing the simulation outputs and the experimental data of the contagion by the Diamond Princess cruise case study (Mizumoto and Chowell, 2020). This environment represents a "close" *world* and, essentially, a closed environment, according to the purposes of the proposed model.

The Diamond Princess cruise[3] is organized in 12 decks open to the passengers, with an overall length of about 290m, and calculated *A* of about 73500m². During the COVID-19 emergency onboard, the cruise host 3711 individuals. The index case was embarked on 20th of January 2020. On the 30th of January, a total of 2 observed cases were reported within the median incubation period (Lauer et al., 2020), and the index case was confirmed on 1st of February 2020. In this study, we consider the increase in observed cases from this day to the maximum daily values in "new" infected people (7th of February), according to the observed data from previous work on the cruise event (Mizumoto and Chowell, 2020). Hence, the overall simulated time is equal to 7 days (420 steps according to Table 2 characterization). This comparison allows calibrating the speed of the virus spreading (analysis of the slope of the increasing part of the contagion curve) in the "close" simulated *world*. The *formcoeff* parameter is used to set up the general model in respect to the calibration reference

---

[2] https://cran.r-project.org/bin/windows/base/; last access: 17/4/2020
[3] all the data are derived by analysing plan schemes reported at https://www.princess.com/ships-and-experience/ships/di-diamond-princess/; last access: 6/4/2020

without changing the general approach described in Section 2.1. Hence, if *formcoeff*=2, the process will ideally be twice faster than for *formcoeff*=1 (default condition in the general model).

The model parameters in the calibration activities are set up by considering:
1. only *formcoeff* as variable, ranging from 0.1 to 2 (by step of 0.1);
2. constant values for the other parameters, as described in Table 2.

About the use-of-space rules, it is assumed that the individuals could spend 15 hours in contact one to each other, according to the general schedule provided by previous works (Fang et al., 2020). Since the model is based on 15minutes-long simulation step, 60 simulation steps are equal to 1 day. No simulation steps are performed during the night-time (hence, no movement simulated). Besides, it is considered that the individuals could freely move from one part of the cruise to another, by moving for about 50m per step, one at one hour.

| Parameter | Value | Source |
|---|---|---|
| A | 73500m$^2$ (271 patches per side) | the overall assessed surface area of the cruise from a graphical evaluation (based on the plans[3]) |
| N | 3711 | (Fang et al., 2020; Mizumoto et al., 2020) |
| $p_{imm}$ | 0 % | no evidence that immune people can exist |
| *Initial infector %* | 0.054% | considering experimental data (Mizumoto and Chowell, 2020), it is calculated as the ratio between 2 observed cases within the median incubation period (Lauer et al., 2020) and N |
| asymptomatic ratio | 20% | the superior limit in the confidence interval of estimated asymptomatic proportion (among all infected cases) (Mizumoto et al., 2020) |
| average delay | 60 | equal to 1 day to be shorter than the time to display symptoms (e.g. fever onset) by the 2.5% of infected persons (Fang et al., 2020; Lauer et al., 2020) |
| *Iinc* | 320 | corresponding to the median incubation time (and the inferior limit of the confidence interval), according to a conservative approach. It corresponds to about 5.1 days (Lauer et al., 2020) |
| *Ifev* | 160 | the average value corresponds to the minimum time to display symptoms by the 2.5% of infected persons (Lauer et al., 2020). A standard deviation is associated to make it ranging from 0 to 320 steps. |
| $prot_i = prot_j$ (*mask filter*) | 0 | It is considered that all the sample is characterized by the same protection level given by the mask filter, to consider uniform conditions in a conservative approach (Fang et al., 2020). No masks are worn in the calibration simulation (Fang et al., 2020; Mizumoto and Chowell, 2020). |
| $t_c$ | 1 step = 15 minutes | consolidated data about closed environment contagion spreading from national and international health organization[1] |
| mask wearing % | 0% | no masks are worn (Fang et al., 2020; Mizumoto and Chowell, 2020) |
| movement at breaks | 20 patches; 50 patches | two supposed distance between the different main locations in the cruise are tested [3] |

*Table 2. Model calibration setup: parameters assumed as constant values.*

In the calibration activities, simulation is run by using the distinct function in the NLRX package, by assigning 100 seeds for each experiment. Results were collected at every 60 steps in order to take into account the daily dynamics of the process.
The number of infected people [pp] per simulation day is evaluated as main comparison output for each sample in the test, and then related average and median values are calculated to be compared to the experimental curve. In particular, the selected *formcoeff* is the one that minimizes the average percentage difference between the infected people assessed by the model median values and the experimental values (Bernardini et al., 2020).

### 2.3. Application to the case study

A part of the main building in the Faculty of Engineering campus at Università Politecnica delle Marche, Ancona (Italy) has been selected as a significant case study. The overall main building hosts about 5000 students and professors, who usually attend lessons in classrooms from Monday to Friday, for an overall lessons time of 8 hours per day. Generally, students are divided into course groups, and each group attend the same lessons in different rooms during the day, thus having the possibility to move from a room to another. Each lesson lasts at least 1 hour. It can be considered that each students' group generally attend lessons in the same area of the building, thus leading to the possibility to have minimum contacts between groups from different areas of the building.

Depending on the case study configuration and on the model calibration activities in Section 2.2, input parameters for the case study simulation are considered as variables (see Table 3) or constant values (see Table 4).

| Parameter | Min | Max | PDFs |
|---|---|---|---|
| N | 250 | 1150 | Uniform |
| Initial infectors % | 0.0546 | 30 | Uniform |
| Mask wearing % | 1 | 100 | Uniform |
| $prot_i = prot_j$ (*mask filter*) | 0 | 1 | Uniform |
| Movement at breaks | 1 | 100 | Uniform |

*Table 3. Parameter characterization for SA analysis*

Table 3 gives an overview of the simulation variables assumed as stochastic parameters, described by Probability Density Functions (PDFs).

In particular, the maximum number of initial people *N* has been defined considering the maximum capacity of the classrooms, which corresponds to about 1pp/m² inside the classroom. The minimum values consider previously available data of occupancy under different scenarios (i.e. lessons, exams, etc…). *Initial infectors %* has been defined by considering as minimum value the input data for the contagion in the Diamond Princess case study (compare to Section 2.2). A maximum value of 30% of the population is arbitrarily chosen to recreate a possible scenario for a "second phase" in the COVID-19 emergency. *Mask wearing %* has been defined as a uniform probability density function. Finally, mask filter has been defined even as a uniform probability density function, considering that the classification of the masks in groups (i.e. FFPX, KXX) is depending on the ability to stop a specific fraction [0-1] of the aerosol drops, which can affect direct and indirect virus spreading in the adopted proximity-based model.

Concerning the constant parameters, the epidemiological variables are scaled according to the calibration model input set, while *formcoeff* is the best-fitting ones according to calibration results from Section 2.2 activities.

Finally, the overall simulated time is equal to 14 days, which corresponds to 10 days of university opening (320 steps according to Table 4 characterization), because this value can represent a good estimation of the maximum incubation time from previous researches (Lauer et al., 2020) and consolidated data from international health organizations[1]. It can be considered as a critical simulation period since the maximum contagion spread due to the initial infected people has been ideally concluded.

The scenario is firstly assumed to perform a Sensitivity analysis, to understand which are the main independent variables affecting the final results (see Section 2.3.1). Then, the same simulation outputs are compared together depending on such SA results (see Section 2.3.2).

| Parameter | Value | Source |
|---|---|---|
| A | 4300m² (66 patches per side) | the overall assessed surface area of the considered part of the building, by including all the spaces accessible by the students (i.e. classrooms, spaces for study) |
| *formcoeff* | Best fitting value according to model calibration activities in Section 2.2 | Deriving the spreading phenomenon according to previous experimental data in a closed environment |
| $p_{imm}$ | 0 % | as for the calibration test, since no evidence that immune people can exist |
| asymptomatic ratio | 20% | as for the calibration test |
| average delay | 32 | equal to 1 day with 8 hours of attendance by students, by scaling the calibration test parameters |
| $I_{inc}$ | 170 | scaling the calibration test parameters depending on the steps per day length |
| $I_{fev}$ | 87 | scaling the calibration test parameters depending on the steps per day length |
| $t_c$ | 1 step = 15 minutes | consolidated data about closed environment contagion spreading from national and international health organization[1] |

*Table 4. Case study application: parameters assumed as constant values.*

### 2.3.1. Sensitivity analysis

A Sensitivity Analysis (SA) has been performed through variance-based decomposition (Sobol′, 2001). The Sobol method is used to calculate, for any stochastic input of the performed calculation, total and first-order sensitivity index (STi, SFi). STi (Sobol total index) measures the contribution to the output variance due to each input, including all variance caused by its interactions with any other input variables (Saltelli et al., 2010, 2008). The higher the value of the sensitivity indices, the most influential the respective input on the outcome. SFi measures indicate the main contribution of each input factor to the variance of the output.

Runs were performed using NLRX package (sobol2007 function) of "R statistics" programming language (Salecker et al., 2019), adopting the Sobol variance decomposition scheme proposed by Saltelli (Saltelli et al., 2010, 2008).

After some preliminary tests necessary to improve the accuracy of the proposed model, we performed two sets of 77000 runs according to the aforementioned parameters setting (see Table 3 and Table 4).

### 2.3.2. Criteria for effectiveness evaluation of mitigation strategies

The results from the simulation scenarios runs performed for Section 2.3.1 are compared together depending on the main independent variables affecting the contagion spreading. Basing on the current solutions in contagion spreading reduction (Fang et al., 2020; Yang et al., 2020; Zhai, 2020), results are discussed in terms of:

1. effect of *mask filter* as individuals' protection solution, combined with the *mask wearing %* (classified in homogeneous classes with steps of 10%), that represents the implementation level for the solution. The multiplication between *mask filter* and *mask wearing %* is introduced to have a quick evaluation index combining these two individuals' protection solution-related parameters;
2. *N* as density related factors which can affect the possibility to implement social distancing solutions; *N* values can be classified according to occupants' density *Docc* [pp/m$^2$] in classrooms, which are the permanence areas for the users. *Docc* values are offered by 0.1pp/m$^2$ wide classes. Discretization in 4 density classes is also provide to discuss the effects in relation to the average dimension of the seats in the classrooms: (a) 250≤*N*≤350, 0.2pp/m$^2$≤*Docc*≤0.3pp/m$^2$; (b) 350<*N*≤600, 0.3pp/m$^2$<*Docc*≤0.5pp/m$^2$; (c) 600<*N*≤1000, 0.5pp/m$^2$<*Docc*≤0.7pp/m$^2$; (d) 1000<*N*≤1150, 0.7pp/m$^2$<*Docc*≤1.0pp/m$^2$.
3. *initial infector %* as access control-related factor at the initial simulation step. It is assumed that a good access control solution will detect at least the 95% of the infectors while entering the building starting from the first simulation step. Hence, it is assumed that the implementation of access control strategies will imply *initial infector %* ≤5%; higher values will correspond to no implementation of access control strategies.

These results are organized together to mainly outline different conditions in building operation issues.

As output value, the final infected people percentage *dI* [%] is assessed according to Equation (5)

$$dI = \left[1 - \frac{S_f}{S_{init}}\right]\% \quad (5)$$

where *Sf* [pp] is the final number of susceptible people (not infected) and *Si* [pp] is the initial number of susceptible people (not initially infected). Its trend expresses the contagion spreading within susceptible individuals. Hence, when *dI* tends to 0, all the individuals tend to be not infected while, when *dI* tends to 100%, all the individuals tend to be infected. Solutions effectiveness increases if *dI* is minimized. *dI* values are evaluated at the final simulation step for each considered conditions in the input values, and the distributions of these values are evidenced in respect to the aforementioned input values combinations, by additionally evidencing distribution percentiles. On this distribution, two acceptability limits for the solution effectiveness are selected:

1. *dI*=5%, which implies that at most the 5% of the population will be affected by the virus. This is a conservative limit for the solution effectiveness estimation;
2. *dI*=25%, which implies that at most the 25% of the population will be affected by the virus, according to a quartile analysis of the sample.

Finally, a model resuming the influence of *mask filter*, *mask wearing %* and *Docc* on acceptability limits given by *dI* is proposed to synthetically trace the main simulation results and give a general outlook of the combination between the strategies. To trace such correlation, additional simulation in the range *Docc*=0 to 0.2pp/m2 are performed to represent all the density conditions (also the value under the minimum *N* value considered as acceptable for the building use).

## 3. Results

### 3.1. Model calibration: results

Figure 2 resumes the trend in the number of infected people in the Diamond Princess scenario simulation, by outlining average simulation data for different *formcoeff* values in respect to the experimental data (Mizumoto and Chowell, 2020).

Infected people are represented by logarithm scale. Data are referred to movement at breaks equal to 20 patches, but data for 50 patches have the same trend. Higher the *formcoeff*, obviously higher the final number of the infected people. Furthermore, *formcoeff* affect the shape of the contagion curve, according to the following general average trends:

- for *formcoeff* up to 0.2 (dotted lines in Figure 2), the contagion is characterized by an initial peak in the contagion, and then it appears stable, thus evidencing that the higher delay increasing can have significant long term effects (at more than 420 simulation ticks);
- for *formcoeff* from 0.3 to 0.4 (dashed line in Figure 2), the contagion curve has a similar increasing trend as for the experimental curve, but the higher delay increasing slows down the final result;
- for *formcoeff* from 0.5 to 1.4 (continuous black lines in Figure 2), the contagion curve has a similar increasing trend as for the experimental curve, but the predicted values are higher than those of the experimental data. In this sense, the best trend seems to be related to *formcoeff*=0.5, which additionally allows maintaining a conservative approach in infected people's estimation since it is slightly over the experimental curve;
- for *formcoeff* from 1.5 to 2 (continuous grey lines in Figure 2), the contagion curves have a higher peak in respect to the experimental curve (e.g. for *formcoeff*=2, the peak appears at about 360 steps). Hence, the contagion speed is higher, and the peak of infected people is placed before the experimental one, thus increasing short-term contagion effects.

Figure 3 resumes the trend of the best *formcoeff* value (0.5) in fitting the experimental curve. In this condition, the experimental data are placed close to the median values of the 0.5-related curve, with an average percentage error in predictions equal to about 4% in respect to the 50th percentile values (12% in absolute error terms)[4]. Possible differences in the contagion spreading curve are due to specific issues in the cruise use by passengers (e.g. infectors could not "die" in the real scenario; differences in motion schedule).

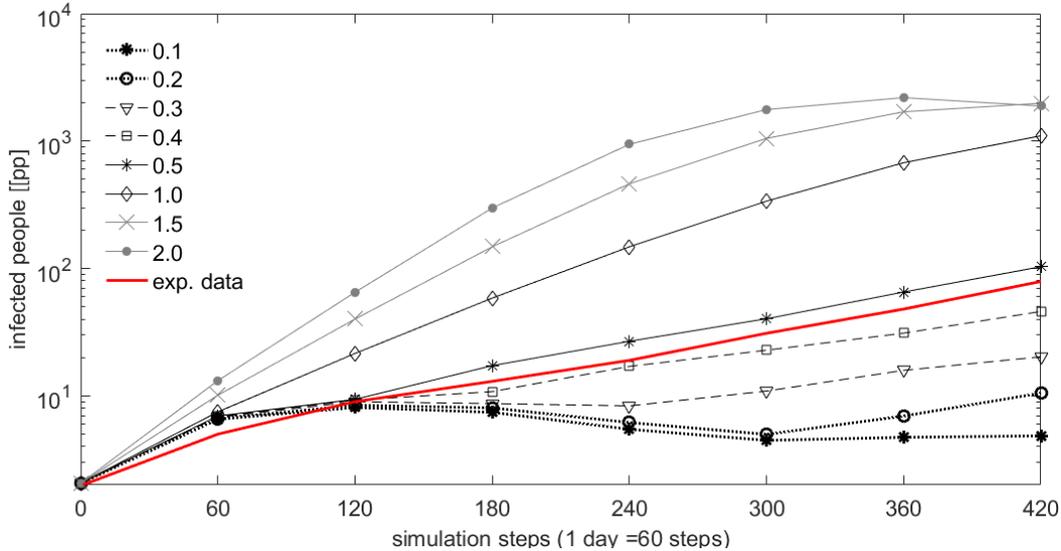

*Figure 2. Comparison between the experimental data from the Diamond Princess cruise (red line) and the simulation results, for the main formcoeff values according to their trend. Infected people are represented by logarithm scale.*

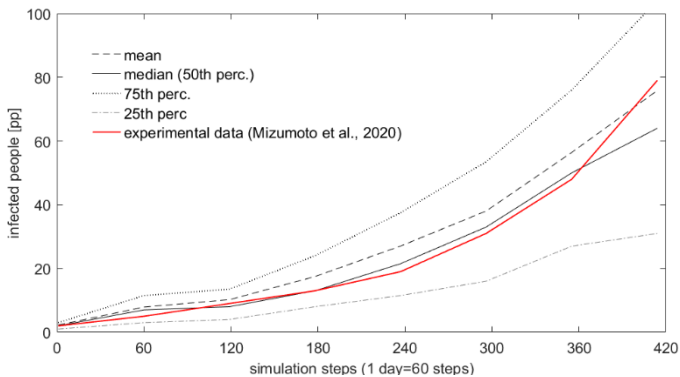

---

[4] tests related to 50 patches offer similar outcomes: average percentage error equal to 11%, and in absolute terms equal to 15%

*Figure 3. Comparison between the experimental data from the Diamond Princess cruise (red line) and the simulation results, for the best formcoeff values (0.5), by considering different percentiles and mean data.*

### 3.2. Application to the case study: results

According to the results in Section 3.1, all the simulations in the case study are performed by considering *formcoeff*=0.5. In the following sections, the SA results are firstly discussed (Section 3.2.1) and the influence of parameters in the case study are offered (Section 3.2.2).

#### 3.2.1. Sensitivity analysis and robustness check

Figure 4 displays the total order sensitivity indices (STi) and the first-order sensitivity indices (SFi).

Considering Total order sensitivity indices (STi), SA suggests that the main source of results' uncertainty is "initial-people" (number of people at the beginning of each simulation run). Considering that spaces where people can move are the same for all the simulations, the value represents the occupation density of the space and also the effect of a possible "social distancing" measure taken to prevent the spread of the contagious disease by maintaining a physical distance between people. The second source of results' uncertainty is *mask filter* (the type of mask adopted in terms of individual protection degree) showing also the importance of individual protection measures. The other secondary sources of results' uncertainty are *Initial infectors %* and *mask wearing %* previously defined. The effect of *movement at breaks* (maximum distance allowed in the movement for each people) appears negligible.

Considering that the sum of first-order sensitivity indices (SFi) is less than 1 (0.94) the model is non-additive, with limited interactions between input factor, as suggested by (Saltelli et al., 2008).

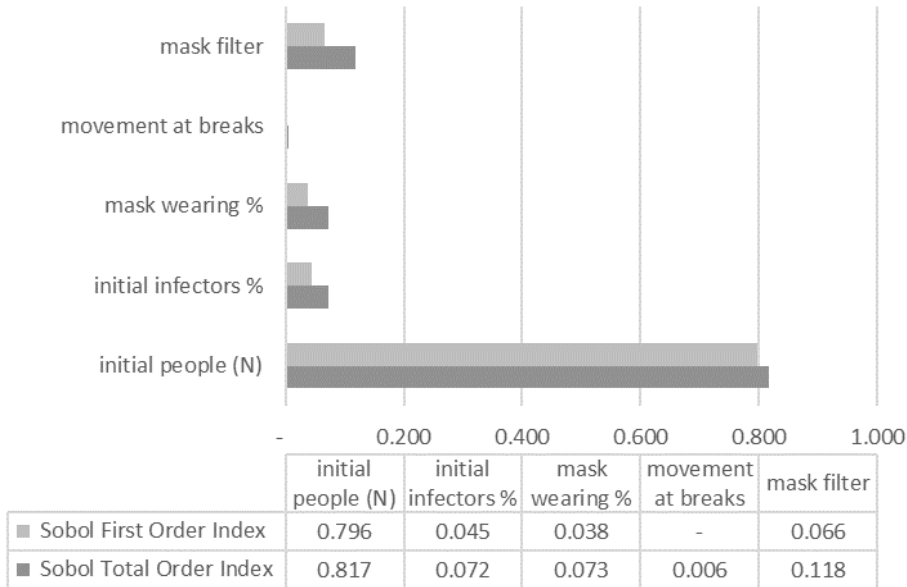

*Figure 4. Total order sensitivity indices (STi – dark grey) and the first-order sensitivity indices (SFi – light grey) for the considered parameters.*

#### 3.2.2. Simulation scenarios results

According to the SA results, the main independent variables are combined to describe different scenarios in the case study.
Hence, the effect of the movement at breaks variable is not here discussed.

##### 3.2.2.1. General influence of the solutions

In general terms, the use of respiratory protective devices with higher *mask filter* values can effectively reduce the virus spreading, especially when the solution is implemented for higher *mask wearing %* values, according to suggestions from previous works (Fang et al., 2020; Zhai, 2020). Figure 9Figure 5 resumes the *dI* distribution according to a boxplot representation (no outliers) for the different *mask filter\*mask wearing%* classes, regardless of density-related and access control-related solutions. Higher *mask filter* values implemented by an increasingly higher number of occupants (*mask wearing %* higher values) imply a reduction in the final number of infected occupants. Acceptability thresholds can be reached when implementing at least *mask filter\*mask wearing% ≥80%* for dI=25% and *≥0.90%* for dI=5%. Such values essentially imply the necessity to implement at least FFP1 masks (or more protective devices) for at least the 90% of the population.

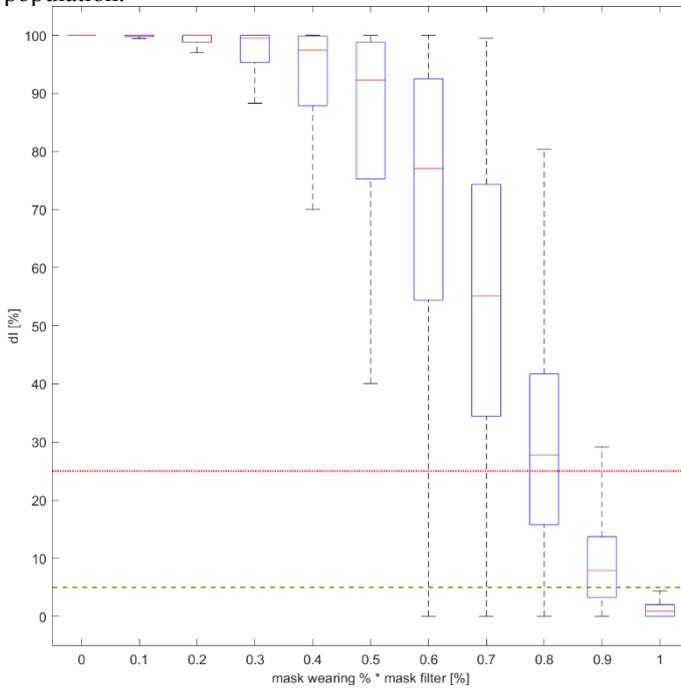

*Figure 5. Boxplot dI values distribution at the last simulation step for the whole sample, with respect to the effects related to mask (mask filter and mask wearing %). dI acceptable thresholds are defined at dI=5% (dashed green line) and 25% (continuous red line).*

*Figure 6* shows how the use of occupants' density control strategies seems to limitedly involve acceptable solutions in terms of *dI* values if applied by themselves. The efficiency of the solution is mainly connected to the possibility to combine such measure to the use of respiratory protective devices, as shown by *Figure 7*. In this sense, the limitation of building use to $D_{occ}$≤0.3pp/m$^2$ (*Figure 7*-A) allows about a 20% reduction in needed *mask filter\*mask wearing%* classes in respect to density maximization conditions (*Figure 7*-D), to obtain dI<25%. Nevertheless, the final result will be affected by the effective possibility to engage users in maintaining social distancing during the whole occupancy timing. This main reason seems to affect the width of boxplot ranges, especially in $D_{occ}$≤0.3pp/m$^2$ regardless of additional solutions (see *Figure 6*) and in intermediate conditions in terms of *mask filter\*mask wearing%* classes in *Figure 7*.

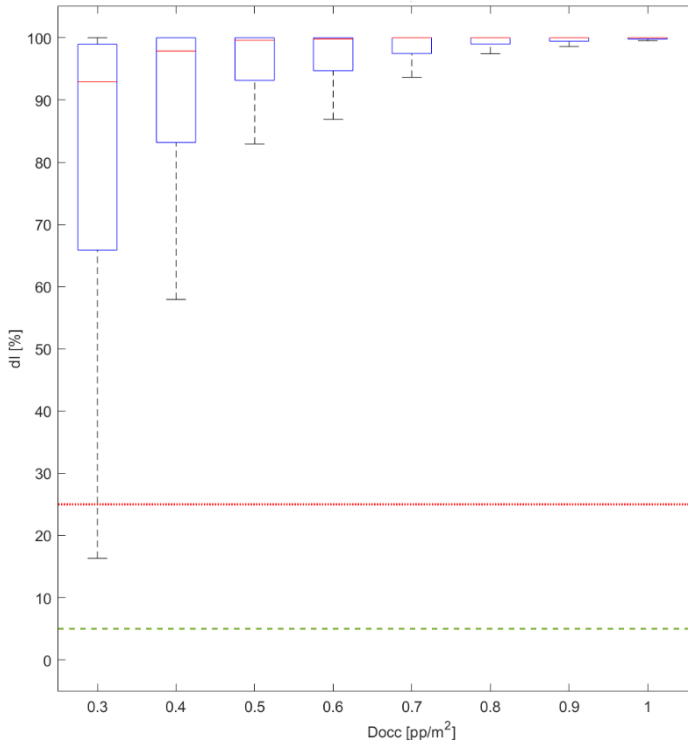

*Figure 6. Boxplot dI values distribution at the last simulation step for the whole sample, with respect to the effect of occupants' density Docc values discretized by 0.1pp/m². dI acceptable thresholds are defined at dI=5% (dashed green line) and 25% (continuous red line).*

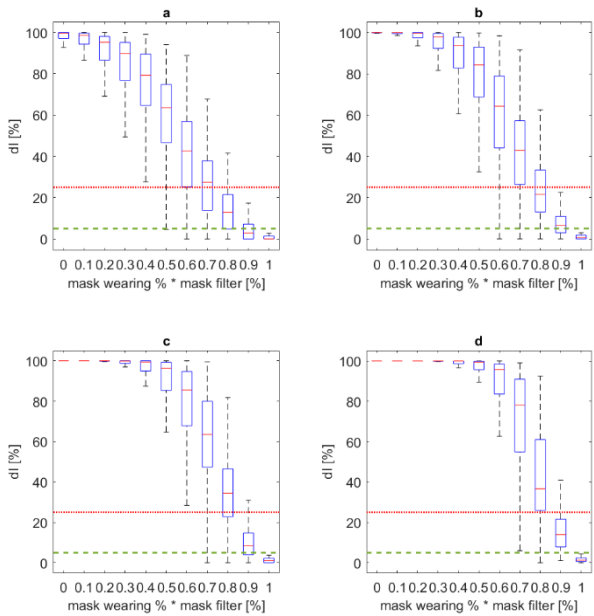

*Figure 7. Boxplot dI values distribution at the last simulation step for the whole sample, with respect to the effects of different density classes: a) $Docc \leq 0.3pp/m^2$; b) $0.3pp/m^2 < Docc \leq 0.5pp/m^2$; c) $0.5pp/m^2 < Docc \leq 0.7pp/m^2$; d) $0.7pp/m^2 < Docc \leq 1.0pp/m^2$. Values are traced according to the overall mask effect. dI acceptable thresholds are defined at dI=5% (dashed green line) and 25% (continuous red line).*

Finally, *Figure 8* suggests how the implementation of access control strategies can significantly improve the effect of respiratory protective devices, especially in the case of more consistent solutions. In this sense, the limitation of building access (*Figure 7*-A) allows about 10% reduction in needed *mask filter\*mask wearing%* classes in respect to conditions in which no access control are performed, to obtain the outcoming median of dI<5%.

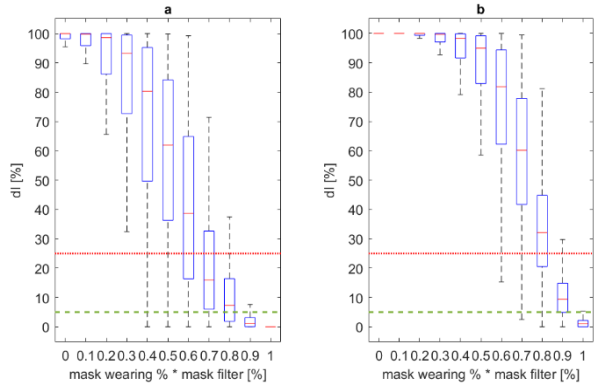

*Figure 8. Boxplot dI values distribution at the last simulation step for the whole sample, with respect to: a) access control strategies implemented and b) no access control strategies implemented. Values are traced according to the overall mask effect. dI acceptable thresholds are defined at dI=5% (dashed green line) and 25% (continuous red line).*

### 3.2.2.2. Influence of mask filter at maximum building capacity

According to the general remarks for the whole sample, it could be possible to face maximum building capacity conditions ($0.7pp/m^2<D_{occ}≤1.0pp/m^2$) by firstly implementing mask filter-based solutions. Figure 9 traces, in the different panels, the effects of mask filter classes on *dI*, depending on the level of implementation within the hosted population (*mask wearing %*), regardless of the access control strategies. The implementation of FFP3 masks for more than 90% of the occupants leads to *dI* values under the *dI* thresholds. The same result is reached in case of no access control strategies implemented, by considering FFP3 and FFP2 implementation, as shown by Figure 10. This outcome is due to the poor effect given by the implementation of access control strategies on the highest mask filter-based solutions, as remarked by *Figure 11*: the main effect given by the implementation of access control strategies is the reduction of the overall dispersion of *dI* data for the class *mask wearing% * mask filter*=80% (see *Figure 11*-A), which is -5% dispersed in respect to the same class for no access control strategies conditions (see *Figure 11*-B), by considering the interval 25th-75th percentiles. Nevertheless, *Figure 12* shows how poorer mask filter based solutions can take advantages of the access control strategies implementation: in particular, using FFP1 by about the 100% of the population in access control conditions could lead at *dI*<25% for more than the 50% of the simulated cases (*Figure 12*-A; here the FFP1 median at *mask wearing%*=100% is equal to about 10%), while extreme cases for surgical mask implementation falls under the limit dI thresholds(*Figure 12*-B).

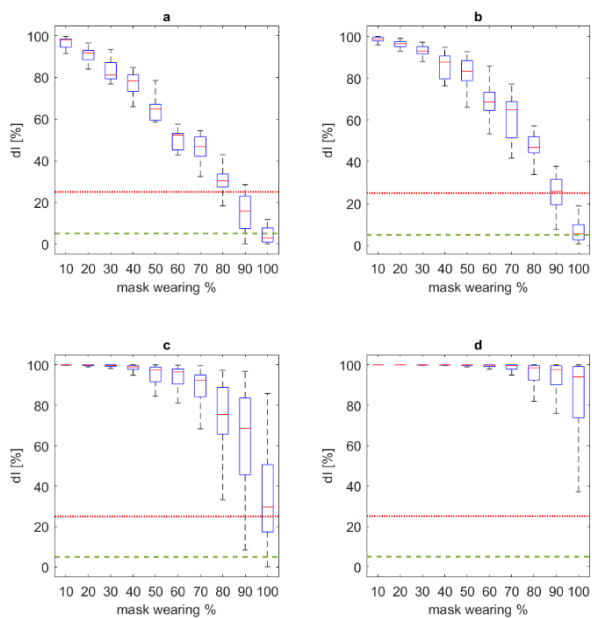

*Figure 9. Boxplot dI values distribution at the last simulation step for maximum building capacity, in respect to the effects of different mask filter classes: a) FFP3; b) FFP2; c) FFP1; d) surgical mask. The Boxplot representation is offered by distinguishing the different mask wearing % classes. dI acceptable thresholds are defined at dI=5% (dashed green line) and 25% (continuous red line).*

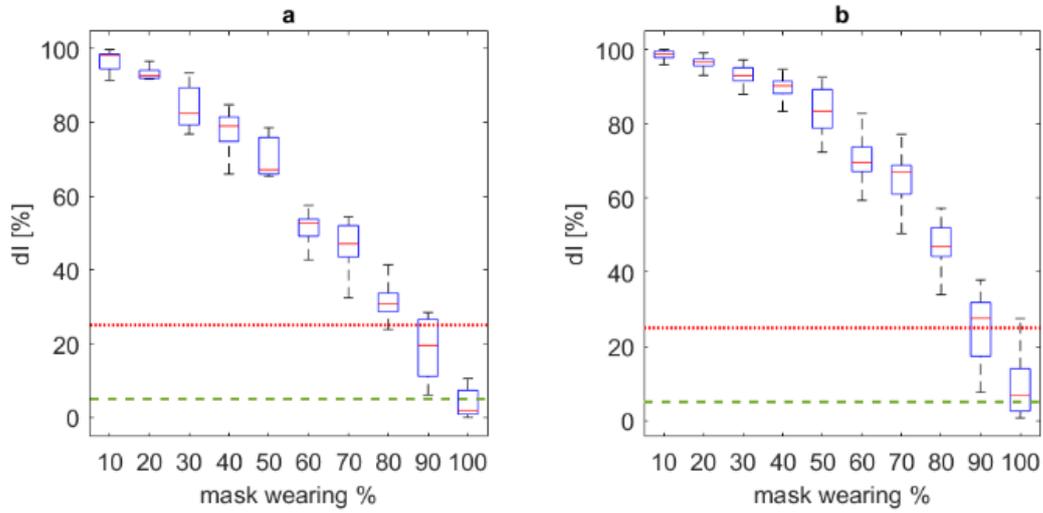

*Figure 10. Boxplot dI values distribution at the last simulation step for maximum building capacity in no access control strategies conditions, in respect to the effects of different mask filter classes: a) FFP3; b) FFP2. The Boxplot representation is offered by distinguishing the different mask wearing % classes. dI acceptable thresholds are defined at dI=5% (dashed green line) and 25% (continuous red line).*

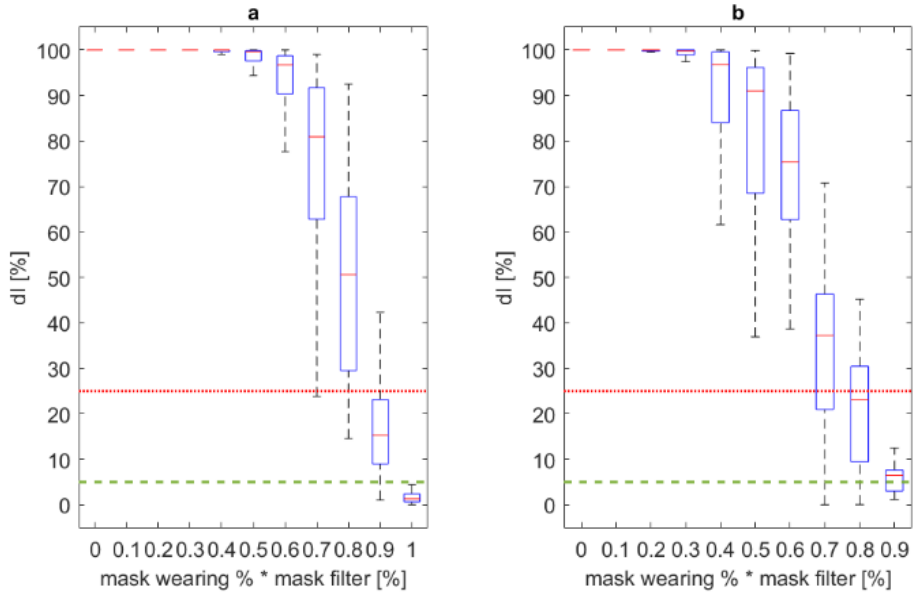

*Figure 11. Boxplot dI values distribution at the last simulation step for maximum building capacity, in respect to: a) access control strategies implemented and b) no access control strategies implemented. Values are traced according to the overall mask effect. dI acceptable thresholds are defined at dI=5% (dashed green line) and 25% (continuous red line).*

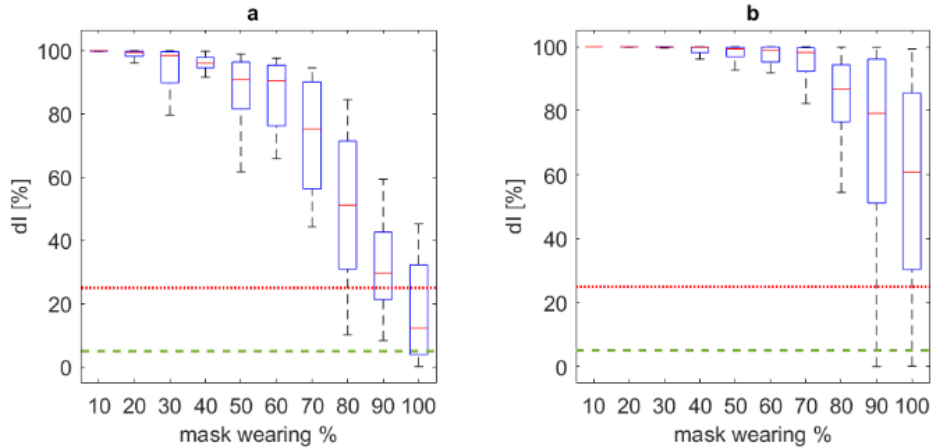

*Figure 12. Boxplot dI values distribution at the last simulation step for maximum building capacity when access control strategies are considered, in respect to the effects of different mask filter classes: a) FFP1; b) surgical masks. The Boxplot representation is offered by distinguishing the different mask wearing % classes. dI acceptable thresholds are defined at dI=5% (dashed green line) and 25% (continuous red line).*

### 3.2.2.3. Influence of occupants' density in poor mask-filter based solutions

Figure 13 and Figure 14 shows dI distribution values by boxplot representation respectively depending on the adoption of surgical masks and non-standards protection solutions by the users, for scenarios where access control strategies are implemented (Figure 13-B and Figure 14-B) or not (Figure 13-A and Figure 14-A), regardless of the *mask wearing %*. These results shows that:
- a minimum protection level in terms of mask filter should be always guaranteed to the occupants to have limit conditions within the *dI* acceptability thresholds;
- access control strategies in such conditions are always recommended and better benefits can be related to the implementation of occupants' density control.

The impact of access control strategies is confirmed by *Figure 15*, which demonstrates the impact of *Docc*-related implementation strategies when surgical masks are used by occupants and access control strategies are maintained. A reduction of about 20% in mask wearing % effort could be achieved by passing from $0.5pp/m^2 < Docc \leq 0.7pp/m^2$ (*Figure 15*-C) to $Docc \leq 0.3pp/m^2$ (*Figure 15*-A), if considering the dI threshold at 25%. Anyway, $Docc > 0.3pp/m^2$ conditions should be avoided, according to *Figure 15*.

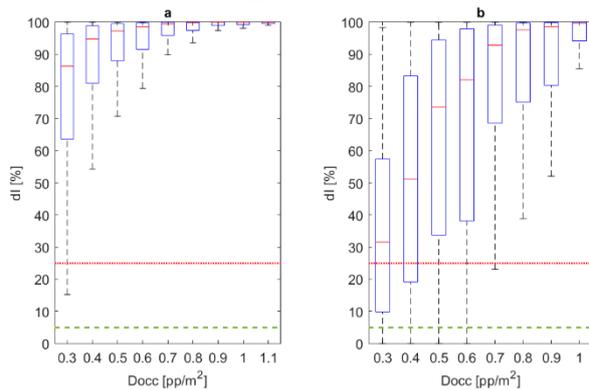

*Figure 13. Boxplot dI values distribution at the last simulation step for surgical masks implementation in relation to the occupants' density Docc classes, with respect to: a) no access control strategies implemented; b) access control strategies implemented. Data are offered regardless of the mask wearing % classes. dI acceptable thresholds are defined at dI=5% (dashed green line) and 25% (continuous red line).*

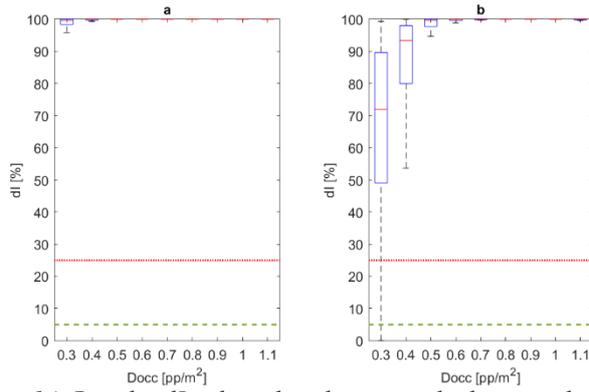

*Figure 14. Boxplot dI values distribution at the last simulation step for non-standards protection (0 to 0.25, compare to Table 1) solutions, in relation to the occupants' density Docc classes, with respect to: a) no access control strategies implemented; b) access control strategies implemented. Data are offered regardless of the mask wearing % classes. dI acceptable thresholds are defined at dI=5% (dashed green line) and 25% (continuous red line).*

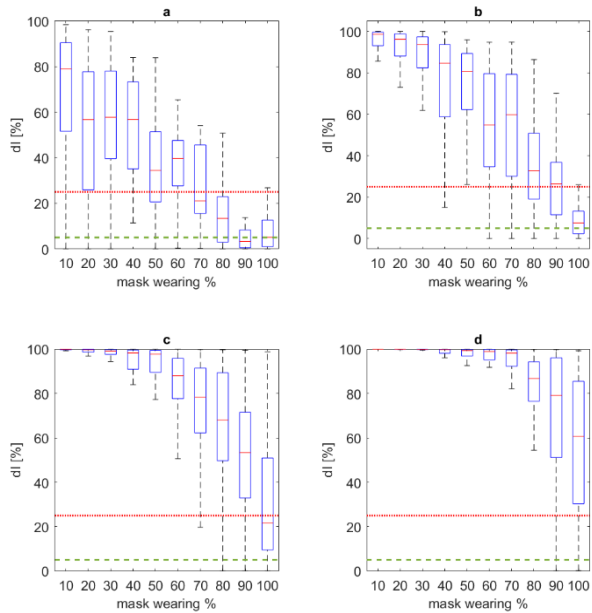

*Figure 15. Boxplot dI values distribution at the last simulation step for surgical mask implementation scenarios, with respect to the effects of different density classes: a) $Docc \leq 0.3 pp/m^2$; b) $0.3 pp/m^2 < Docc \leq 0.5 pp/m^2$; c) $0.5 pp/m^2 < Docc \leq 0.7 pp/m^2$; d) $0.7 pp/m^2 < Docc \leq 1.0 pp/m^2$. Values are traced according to the overall mask effect. dI acceptable thresholds are defined at dI=5% (dashed green line) and 25% (continuous red line).*

## 4. Discussion in the view of restart in public buildings and of modelling tools application

Results show that the effective possibility to limit the virus spreading while restarting activities in public buildings during the "second phase" of the COVID19 emergency could be possible only if more than 1 risk-reduction solution will be implemented.

In general terms, the simulation outcomes evidence how using respiratory protective devices by occupants seems to generally lead to safer conditions for the occupants, especially if high protection measures will be adopted in terms of mask filter ("invasive" masks, like FFP ones) and of widespread implementation (about all the occupants should wear them). The reduction in the building occupants' density could reduce the needed effort in terms of respiratory protective devices especially when access control strategies are implemented to reduce the initial number of infectors entering the building. In this term, the simulation results agree with previous preliminary insights on the importance of facial masks to limit the contagion spreading after the activities restarting (Howard et al., 2020; Zhai, 2020). The application of such strategy will lead to move towards the adoption of less invasive masks, e.g. surgical mask, by the whole number of the hosted individuals, as also suggested by review studies (Howard et al., 2020). Such a solution could be more acceptable by the final users.

From a stakeholder's perspective, the sustainability level depends on the specific aspects involving the strategies:
1. for mask implementation strategies: ensuring that the building users should wear facial masks characterized by a specified *mask filter* for at least the considered *mask wearing %* will imply economic (i.e. costs for masks) and operational (i.e. activities for distribution of masks to the occupants) evaluations;
2. for density control strategies: the possibility to continue the building activities with a reduced number of users (e.g. limiting the number of occupants by guaranteeing "remote access" to all the others) should be considered depending on the activities hosted in the building. This aspect should be combined to economic evaluations as well as to the possibility to effectively deploy the measures into the building (e.g. technological access control implementation; use of building staff members at the building entrances; occupants' positioning control solutions), also in relation to facial masks use to increase the occupants' density;
3. for access control strategies: deploying staff members or technological solutions to check the users' health state at the entrance should face both economic and operational issues, but also the possibility to guarantee rapid access by the users themselves.

Results could be also used to derive simplified rules to combine the strategies together to reach, at most, the acceptable *dI* value. *Figure 16* traces the correlation between *mask filter* values and *Docc* that lead to final simulation results with *dI* placed under the considered acceptability threshold (5% for *Figure 16*-A; 25% for *Figure 16*-B), regardless of the *initial infectors %*. The mask filter classes adopted in previous section 3.2.2 are also shown, while the colour of data refers to the related mask wearing *%* values (colour bar on the right). Finally, the interpolation of maximum values is offered according to a power-based regression approach ($ax^b+c$). It is better to evidence that the provided equations just try to give a first rough quantitative measure of the upper boundary limit not to be crossed. This means that no admitted solutions are present over the curve.

The regression shows the limit values for *mask filter values - Docc* pairs which can lead to acceptable scenarios according to the defined threshold. Lower values of *mask filter* imply lower maximum densities, also according to Section 3.2.2.3 outcomes. As previously pointed out by Section 3.2.2.2, *Docc* equal to about 1pp/m$^2$ are only admitted for FFP3 masks implementation.

As expected, maximum pairs and the related regression curves at *dI*=5% are lower than those at *dI*=25%. The general trend and the regression coefficient data confirms simulation results. Some extreme cases could be highlighted by the regression trend. Considering minimum *mask filter* value, *Docc* tends to 0.3pp/m$^2$ for dI=25%: this density condition essentially refers to the lower cases in the boxplot of *Figure 14*-B, which refers to the implementation of non-standards protection solutions in combination with access control strategies. Regression predictions move closer for FFP2 and FFP3 mask implementation. Finally, the 95% of confidence intervals regression curves evidence possible upper and lower bounds of the models. In case of a more conservative approach, it should better use the lowest dashed curves (lower confidence bond at 2.5%), which essentially admit as an acceptable solution the adoption of surgical masks at the lowest densities (e.g.: for dI=5%: for Docc ranging from 0.0 to 0.1pp/m$^2$, the *mask filter* is close to 0.54).

Since the model does not include the *mask wearing %* as prediction input, the proposed regression model can be adopted also when the users wear respiratory protective devices with different *mask filter* values, by conservatively considering that the *mask filter* to be considered will be the one of the lower one. Furthermore, the *mask wearing %* data on *Figure 16* outline the minimum values for the pair implementation.

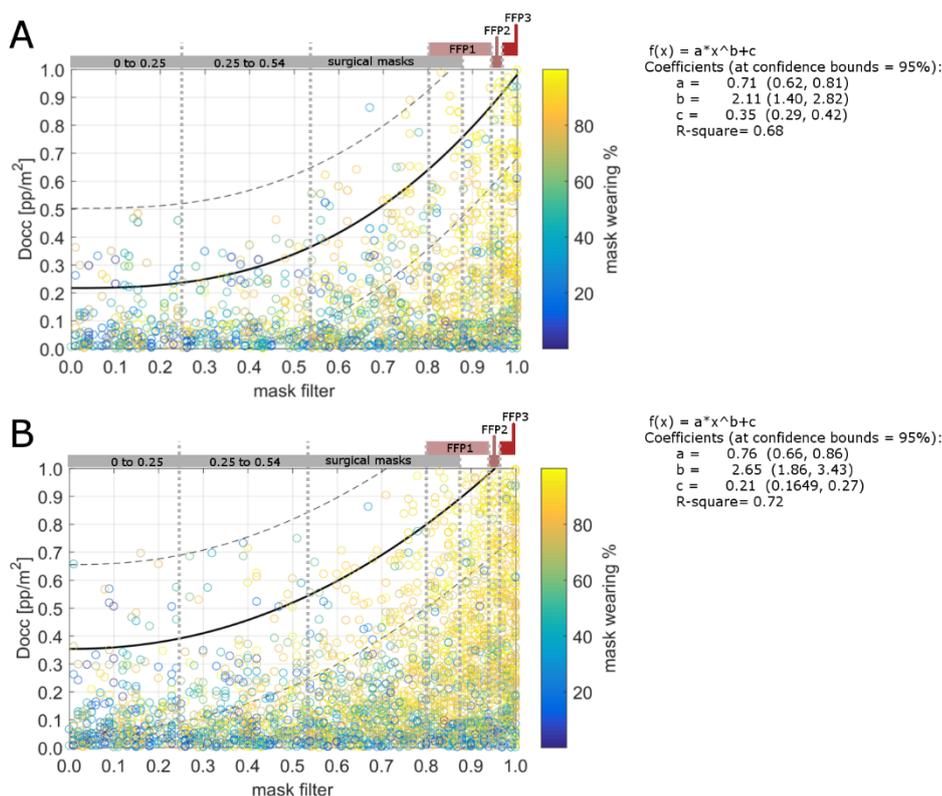

*Figure 16. Mask filter-Docc correlation for all the pairs related to a) dI≤5% and b) dI≤25%. The pairs' colour is related to the mask wearing %. Regression curves ($ax^b+c$) are shown by providing 95% of confidence intervals regression (dashed lines; see equation coefficients on the right of each panel). Mask filter classes are also shown in the upper part of the graphs.*

From a modelling perspective, results confirm how the developed simulator could effectively represent different scenarios in buildings depending on the input factors used to estimate the virus spreading probability. The calibration of the model is performed from current available experimental data, to reproduce real-world conditions in the simulated environment. When other data will be available, it will be possible to update such calibration, by also including, for example, additional modes of person-to-person transmission means or person-to-building components (i.e., contact with contaminated objects) or systems (e.g. ventilation systems) (Adams et al., 2016; Emmerich et al., 2013; Gao et al., 2016; Noakes and Sleigh, 2009; Pica and Bouvier, 2012; Zhang et al., 2018).

Besides, specification on the layout in terms of geometry and presence of obstacles could be implemented, to better take into account the additional architectural features affecting the virus spreading according to consolidate literature on airborne diseases (Pica and Bouvier, 2012; Prussin et al., 2020).

Finally, the model considers the possibility to have "new" infected individuals only inside the building. This aspect could effectively represent most of the basic "second phase" conditions in which the agents' life does not include the participation to social activities and limit "unsafe" contacts outside of the "working" (or "studying") places. Anyway, the daily infector percentage in the model could vary if considering additional new infected people/infectors whose contagion happened outside of the building.

## 4. Conclusions

In view of the restart of public buildings and their activities after the lockdown phase for the COVID-19 emergency, mitigation strategies should be implemented in the closed built environment to avoid secondary peaks in the virus spreading among the communities. In this sense, simulation tools could support the stakeholders' decisions in the view of optimizing the measures to be adopted.

This paper provides the development of an agent-based model to estimate the virus spreading in closed environments by considering a proximity-based approach, which refers to consolidated analysis on virus transmission modes. The proposed model can be applied to public buildings hosting i.e. offices, university or schools, which are characterized by the same daily users. The model can include the representation of contagion-mitigation strategies connected to the use of facial masks/respiratory protective devices, occupants' density control (in the view of maintaining social-distancing strategies) and other access control strategies. It is also able to provide a first quantitative measure of the implementation of such strategies by the relative stakeholders. After its implementation in a software tool and the calibration activities (through available experimental data in a closed environment), the simulator is applied to a significant case study (university building).

Results show how the model seems to faithfully reproduce what happened in closed environments, thus being able to represent the impact of the contagion-mitigation strategies to limit the contagion spreading during the time, i.e. the use of facial masks by the users, the limitation of occupants' densities. Hence, the effectiveness of each strategy (as individually adopted) and of the combination between different measures could be evaluated by the simulator.

In the considered case study, results show that the major effectiveness seems to be reachable by using respiratory protective devices (i.e.e FFPk mask), regardless of the implementation of other strategies. Nevertheless, the adoption of FFPk devices can generally imply a low level of acceptability by the users, since they are not comfortable and easy-to-use by non-specialized users. An acceptable level of effectiveness could be reached by combining different measures, i.e. the use of facial masks to the control of occupants' densities. Limiting the number of people inside the building could support the implementation of surgical masks by the users, thus improving the mask-related operational conditions for the occupants. According to the results for the case study application evidence, the occupants' density for users wearing surgical masks should be at most halved to reach the same effectiveness of full occupants' density with FFP3 mask worn. The stakeholders could support operational decisions by the model, according to both cost-benefits and users' acceptability standpoints.

Finally, thanks to the agent-based approach, the model could be easily modified to integrate different epidemiological data (including modes of virus transmission), different build environment (e.g. to include layout features) and rules for spaces use by occupants, as well as to apply it to other relevant contexts for the "second phase" restart, such as tourist facilities, cultural buildings and so on.

## 5. References

Adams, R.I., Bhangar, S., Dannemiller, K.C., Eisen, J.A., Fierer, N., Gilbert, J.A., Green, J.L., Marr, L.C., Miller, S.L., Siegel, J.A., Stephens, B., Waring, M.S., Bibby, K., 2016. Ten questions concerning the microbiomes of buildings. Building and Environment 109, 224–234. doi:10.1016/j.buildenv.2016.09.001

Banos, A., Lang, C., Marilleau, N., 2015. Agent-Based Spatial Simulation with NetLogo, Agent-Based Spatial Simulation with NetLogo. doi:10.1016/c2015-0-01299-0

Barbieri, C., Darnis, J.-P., 2020. Technology: An Exit Strategy for COVID-19? IAI COMMENTARIES 20, 1–6.

Bernardini, G., Postacchini, M., Quagliarini, E., Brocchini, M., Cianca, C., D'Orazio, M., 2017. A preliminary combined simulation tool for the risk assessment of pedestrians' flood-induced evacuation. Environmental Modelling & Software 96, 14–29. doi:10.1016/j.envsoft.2017.06.007

Bernardini, G., Quagliarini, E., D'Orazio, M., Brocchini, M., 2020. Towards the simulation of flood evacuation in urban scenarios: Experiments to estimate human motion speed in floodwaters. Safety Science 123, 104563. doi:10.1016/j.ssci.2019.104563

Bourouiba, L., 2020. Turbulent Gas Clouds and Respiratory Pathogen Emissions. JAMA. doi:10.1001/jama.2020.4756

Chan, M., Campo, E., Estève, D., Fourniols, J.-Y., 2009. Smart homes - current features and future perspectives. Maturitas 64, 90–7. doi:10.1016/j.maturitas.2009.07.014

Chen, W., Zhang, N., Wei, J., Yen, H.-L., Li, Y., 2020. Short-range airborne route dominates exposure of respiratory infection during close contact. Building and Environment 176, 106859. doi:10.1016/j.buildenv.2020.106859

D'Orazio, M., Quagliarini, E., Bernardini, G., Spalazzi, L., 2014. EPES - Earthquake pedestrians' evacuation simulator: A tool for predicting earthquake pedestrians' evacuation in urban outdoor scenarios. International Journal of Disaster Risk Reduction 10, 153–177. doi:10.1016/j.ijdrr.2014.08.002

Emmerich, S.J., Heinzerling, D., Choi, J., Persily, A.K., 2013. Multizone modeling of strategies to reduce the spread of airborne infectious agents in healthcare facilities. Building and Environment 60, 105–115. doi:10.1016/j.buildenv.2012.11.013

Face ID firms battle Covid-19 as users shun fingerprinting, 2020. . Biometric Technology Today 2020, 1–2. doi:10.1016/S0969-4765(20)30042-4

Fanelli, D., Piazza, F., 2020. Analysis and forecast of COVID-19 spreading in China, Italy and France. Chaos, Solitons & Fractals 134, 109761. doi:10.1016/j.chaos.2020.109761

Fang, Z., Huang, Z., Li, X., Zhang, J., Lv, W., Zhuang, L., Xu, X., Huang, N., 2020. How many infections of COVID-19

there will be in the "Diamond Princess"-Predicted by a virus transmission model based on the simulation of crowd flow.

Gao, X., Wei, J., Lei, H., Xu, P., Cowling, B.J., Li, Y., 2016. Building Ventilation as an Effective Disease Intervention Strategy in a Dense Indoor Contact Network in an Ideal City. PLOS ONE 11, e0162481. doi:10.1371/journal.pone.0162481

Howard, J., Huang, A., Li, Z., Tufekci, Z., Zdimal, V., van der Westhuizen, H., von Delft, A., Price, A., Fridman, L., Tang, L., Tang, V., Watson, G.L., Bax, C.E., Shaikh, R., Questier, F., Hernandez, D., Chu, L.F., Ramirez, C.M., Rimoin, A.W., 2020. Face Masks Against COVID-19: An Evidence Review. Preprints 2020040203. doi:10.20944/preprints202004.0203.v1

Laskowski, M., Demianyk, B.C.P., Witt, J., Mukhi, S.N., Friesen, M.R., McLeod, R.D., 2011. Agent-Based Modeling of the Spread of Influenza-Like Illness in an Emergency Department: A Simulation Study. IEEE Transactions on Information Technology in Biomedicine 15, 877–889. doi:10.1109/TITB.2011.2163414

Lauer, S.A., Grantz, K.H., Bi, Q., Jones, F.K., Zheng, Q., Meredith, H.R., Azman, A.S., Reich, N.G., Lessler, J., 2020. The Incubation Period of Coronavirus Disease 2019 (COVID-19) From Publicly Reported Confirmed Cases: Estimation and Application. Annals of Internal Medicine. doi:10.7326/M20-0504

Lopez, L.R., Rodo, X., 2020. A modified SEIR model to predict the COVID-19 outbreak in Spain and Italy: simulating control scenarios and multi-scale epidemics. (preprint on www.medrxiv.org). doi:10.1101/2020.03.27.20045005

Mizumoto, K., Chowell, G., 2020. Transmission potential of the novel coronavirus (COVID-19) onboard the diamond Princess Cruises Ship, 2020. Infectious Disease Modelling 5, 264–270. doi:10.1016/j.idm.2020.02.003

Mizumoto, K., Kagaya, K., Zarebski, A., Chowell, G., 2020. Estimating the asymptomatic proportion of coronavirus disease 2019 (COVID-19) cases on board the Diamond Princess cruise ship, Yokohama, Japan, 2020. Eurosurveillance 25. doi:10.2807/1560-7917.ES.2020.25.10.2000180

Murray, O.M., Bisset, J.M., Gilligan, P.J., Hannan, M.M., Murray, J.G., 2020. Respirators and surgical facemasks for COVID-19: implications for MRI. Clinical Radiology. doi:10.1016/j.crad.2020.03.029

Noakes, C.J., Sleigh, P.A., 2009. Mathematical models for assessing the role of airflow on the risk of airborne infection in hospital wards. Journal of The Royal Society Interface 6. doi:10.1098/rsif.2009.0305.focus

Pica, N., Bouvier, N.M., 2012. Environmental factors affecting the transmission of respiratory viruses. Current Opinion in Virology 2, 90–95. doi:10.1016/j.coviro.2011.12.003

Prem, K., Liu, Y., Russell, T.W., Kucharski, A.J., Eggo, R.M., Davies, N., Jit, M., Klepac, P., Flasche, S., Clifford, S., Pearson, C.A.B., Munday, J.D., Abbott, S., Gibbs, H., Rosello, A., Quilty, B.J., Jombart, T., Sun, F., Diamond, C., Gimma, A., van Zandvoort, K., Funk, S., Jarvis, C.I., Edmunds, W.J., Bosse, N.I., Hellewell, J., 2020. The effect of control strategies to reduce social mixing on outcomes of the COVID-19 epidemic in Wuhan, China: a modelling study. The Lancet Public Health. doi:10.1016/S2468-2667(20)30073-6

Prussin, A.J., Belser, J.A., Bischoff, W., Kelley, S.T., Lin, K., Lindsley, W.G., Nshimyimana, J.P., Schuit, M., Wu, Z., Bibby, K., Marr, L.C., 2020. Viruses in the Built Environment (VIBE) meeting report. Microbiome 8, 1. doi:10.1186/s40168-019-0777-4

Rengasamy, S., Shaffer, R., Williams, B., Smit, S., 2017. A comparison of facemask and respirator filtration test methods. Journal of Occupational and Environmental Hygiene 14, 92–103. doi:10.1080/15459624.2016.1225157

Saari, A., Tissari, T., Valkama, E., Seppänen, O., 2006. The effect of a redesigned floor plan, occupant density and the quality of indoor climate on the cost of space, productivity and sick leave in an office building–A case study. Building and Environment 41, 1961–1972. doi:10.1016/j.buildenv.2005.07.012

Salecker, J., Sciaini, M., Meyer, K.M., Wiegand, K., 2019. The nlrx r package: A next-generation framework for reproducible NetLogo model analyses. Methods in Ecology and Evolution 10, 1854–1863. doi:10.1111/2041-210X.13286

Saltelli, A., Annoni, P., Azzini, I., Campolongo, F., Ratto, M., Tarantola, S., 2010. Variance based sensitivity analysis of model output. Design and estimator for the total sensitivity index. Computer Physics Communications 181, 259–270. doi:10.1016/j.cpc.2009.09.018

Saltelli, A., Ratto, M., Andres, T., Campolongo, F., Cariboni, J., Gatelli, D., Saisana, M., Tarantola, S., 2008. Global Sensitivity Analysis. The Primer, Global Sensitivity Analysis. The Primer. doi:10.1002/9780470725184

Servick, K., 2020. Cellphone tracking could help stem the spread of coronavirus. Is privacy the price? Science. doi:10.1126/science.abb8296

Sobol′, I.., 2001. Global sensitivity indices for nonlinear mathematical models and their Monte Carlo estimates. Mathematics and Computers in Simulation 55, 271–280. doi:10.1016/S0378-4754(00)00270-6

Wilder-Smith, A., Chiew, C.J., Lee, V.J., 2020. Can we contain the COVID-19 outbreak with the same measures as for SARS? The Lancet Infectious Diseases. doi:10.1016/S1473-3099(20)30129-8

Wilensky, U., 1999. NetLogo. http://ccl.northwestern.edu/netlogo/ [WWW Document]. Center for Connected Learning and Computer-Based Modeling, Northwestern University, Evanston, IL.

Yang, Y., Peng, F., Wang, R., Guan, K., Jiang, T., Xu, G., Sun, J., Chang, C., 2020. The deadly coronaviruses: The 2003 SARS pandemic and the 2020 novel coronavirus epidemic in China. Journal of Autoimmunity 102434. doi:10.1016/j.jaut.2020.102434

Zhai, Z., 2020. Facial mask: A necessity to beat COVID-19. Building and Environment 175, 106827. doi:10.1016/j.buildenv.2020.106827

Zhang, N., Huang, H., Su, B., Ma, X., Li, Y., 2018. A human behavior integrated hierarchical model of airborne disease transmission in a large city. Building and Environment 127, 211–220. doi:10.1016/j.buildenv.2017.11.011

Zheng, X., Zhong, T., Liu, M., 2009. Modeling crowd evacuation of a building based on seven methodological approaches. Building and Environment 44, 437–445. doi:10.1016/j.buildenv.2008.04.002